\documentstyle[editedvolume]{crckapb} 

\newcommand{\hb}{\hfill\break}
\input epsf
\def\spose#1{\hbox to 0pt{#1\hss}}
\def\lta{\mathrel{\spose{\lower 3pt\hbox{$\mathchar"218$}}
     \raise 2.0pt\hbox{$\mathchar"13C$}}}
\def\gta{\mathrel{\spose{\lower 3pt\hbox{$\mathchar"218$}}
     \raise 2.0pt\hbox{$\mathchar"13E$}}}
\def\etal{{\it et al.\ }}
\begin{opening}
\title{Chemical Evolution of Spiral Galaxies \protect\\
from Redshift 4 to the Present}
\author{U. Fritze - v. Alvensleben, U. Lindner, K. J. Fricke}
\institute{Universit\"ats--Sternwarte, G\"ottingen, Germany}
\end{opening}
\runningtitle{Chemical Evolution of Spiral Galaxies}
\begin{document}

\section{Introduction}
ISM abundances in nearby spiral galaxies are well 
known from HII region studies (Zaritsky \etal 1994). While 
early type spirals, Sa, Sb, have rather uniform abundances and a narrow range of present 
star formation rates ({\bf SFR}) the galaxy-to-galaxy variations both in HII 
region abundances and in present SFR increase towards late spiral types Sc, Sd 
(see e.g. Kennicutt \& Kent 1983). 
ISM abundances of spiral galaxies or their progenitors up to the 
highest redshifts can be studied 
via the absorption properties imprinted in the spectra of background QSOs. 
While MgII- and CIV- absorption lines are produced in the low column density gas 
of the extended haloes around galaxies, the Damped Ly$\alpha$ 
Absorption ({\bf DLA}) is believed to originate in (proto-)galactic disks. 
High resolution spectroscopy of a large number of metal lines associated with 
DLA systems reveal the redshift evolution of ISM abundances from 
${\rm z \gta 4}$ to ${\rm z \sim 0.6}$. \hb
We present a chemo-cosmological evolution model to explore a possible link 
between high redshift DLA systems and nearby spirals. Inclusion of a 
spectrophotometric description in parallel to the chemical modelling 
together with the enormous time baseline corresponding to the redshift 
range until ${\rm z \sim 4}$ allow the 
galaxy parameters to be seriously constrained.

\section{Galaxy Models}
Any kind of evolutionary galaxy model requires two fundamental parameters: the 
star formation history ({\bf SFH}) and the initial mass function ({\bf IMF}). 
For a given set of input physics -- stellar evolutionary tracks, yields, 
spectra, ... -- these two parameters govern both the spectral and the chemical 
evolution of a model galaxy. \hb
Our unified chemical and spectrophotometric galaxy evolution model describes 
in detail the spectrophotometric evolution in terms of luminosities 
${\rm L_U, \dots, L_K}$, colours, gaseous emission lines, 
stellar absorption indices, and the chemical evolution in terms of ISM 
abundances of individual elements 
${\rm ^{12}C}$, ... , ${\rm ^{56}Fe}$, 
including SNI contributions from carbon deflagration white dwarf binaries 
\'a la Matteucci (1991). 
All these quantities are obtained as a function of time, and, for a given 
cosmological model -- characterised by ${\rm H_0, \Omega_0, \Lambda_0}$ and 
a galaxy formation redshift ${\rm z_{form}}$ -- in terms of redshift z. 

\smallskip\noindent
A specific virtue of this kind of evolutionary model is its analytic power. 
Luminosities at any wavelength and the enrichment of any element 
are readily decomposed into contributions from different stellar masses, 
spectral types, 
luminosity classes, 
metallicity subpopulations, 
nucleosynthetic sites (PNe, SNI, SNII, single stars, binaries, ...), and 
all this can be followed as a function of time or redshift.
It's a simplified 1 -- zone model without any dynamics or spatial resolution but 
it consistently accounts for the finite lifetime of each star, i.e. it does not use an 
Instantaneous Recycling Approximation. \hb
To cope with the low metallicities of DLAs we have recently extended our code 
to consistently account for the increasing metallicities of successive generations 
of stars by using sets of input physics (stellar tracks, yields, spectra, ...) 
for a range of metallicities from Z $=0$ to ${\rm Z=2 \cdot Z_{\odot}}$. 
Sources for the stellar yields for various metallicities are v. d. Hoek \& 
Groenewegen (1997) for stars with masses ${\rm\ m_{\ast} \leq 8~M_{\odot}}$ and 
Woosley \& Weaver (1995) for stars in the range ${\rm 12 \leq m_{\ast} \leq 
40~M_{\odot}}$. \hb
We use different SFHs for the different galaxy types in order to 
assure detailed agreement of our model galaxy spectra with template 
spectra for the respective spectral class from Kennicutt (1992). 
While ellipticals 
and haloes are well described with a SFR exponentially decreasing in time, spirals 
are more adequately described by SFRs that are linear functions of 
the evolving gas content G(t). Characteristic timescales for SF 
${\rm t_{\ast}}$, as defined 
by the decrease rate of the initial gas content G$_0$:  
${\rm \int_0^{t_{\ast}} \Psi \cdot dt = 0.63 \cdot G_0}$ are 
${\rm t_{\ast} \sim 
1 ~ Gyr ~(Es ~\&~ Halos), ~t_{\ast} \sim 2 ~Gyr ~(Sa), ~ \sim 3 ~Gyr ~
(Sb), ~\sim 10 ~Gyr ~(Sc),}$ \hfill\break
${\rm  ~ \sim 16 ~Gyr ~(Sd)}$. \hb
While spectrophotometric modelling in comparison to observed galaxy spectra and 
colours allows for reasonable determinations of the SFHs of galaxies 
of various spectral types it gives only weak constraints on the IMF. 
Detailed chemical modelling, on the other hand, constrains both the SFHs 
{\bf and} the IMF through comparison with observations of the redshift 
evolution of abundances and 
abundance ratios of elements originating from different nucleosynthetic 
sites, as e.g. [C/O], [O/Fe], or [Mg/Fe], for an enormous time baseline. 
The  lookback time to redshifts ${\rm z \sim 4}$ corresponds to $\gta 90 \%$ 
of the Hubble time.  \hb
Fritze - v. Alvensleben \etal (1989, 1991) model the redshift evolution of halo 
abundances and compare with MgII- and CIV- data. HST observations of CIV at low redshift 
(Bahcall \etal 1993) confirm our prediction for carbon. 

%

\section{Redshift Evolution of DLA Abundances}
Damped Ly$\alpha$ Absorbers usually feature a series of associated low 
ionisation absorption lines, e.g. of C, N, O, Al, Si, S, Cr, Mn, Fe, Ni, 
Zn, ... . \hb
High resolution spectra allow to derive precise element abundances, which are, 
by now, available for a large 
number of DLAs over the redshift range from z $\sim 0.6$ through z $\gta 4.4$ 
(Lu \etal 1996, Pettini \etal 1997). Considerable care is taken 
to resolve the velocity structure, to only use non- (or weakly) saturated 
lines and dominant ionisation stages, and to a possible 
depletion onto dust grains that may vary dramatically from $\sim 0~ \%$ (for Zn) 
to $\gta 90~ \%$ (for Cr).
We have compiled all available DLA abundances and have carefully 
referred them all to one ``standard'' set of oscillator strengths and 
solar reference values.
Observational abundances of DLAs show a weak increase with decreasing redshift -- 
in particular if compared to CIV - halo - systems -- and a large scatter at 
any redshift. 

\section{Comparison of DLA Abundances with Models for ${\rm\bf Z_{\odot}}$}

For our modelling approach it would not matter if indeed they highest 
redshift DLA systems were a bunch of subgalactic fragments rather than  
one single protodisk. As long as the ensemble of fragments is bound 
to merge into the counterpart of a present-day galactic disk our model SFH is 
then meant to describe the SFH of the ensemble. \hb
Using solar metallicity stellar yields, lifetimes and evolutionary tracks 
as input physics for our modelling, a Scalo IMF, and a ``standard cosmology'' 
(${\rm H_0=50,\Omega_0=1,\Lambda_0=0,z_{form}=5}$), we find the following 
results (Fig. 1): 
\begin{itemize}
\item[$\bullet$] Sa -- Sd models nicely bracket the DLA abundance data, i.e., 
\item[$\bullet$] for SF timescales  $t_{\ast}$ typical of spiral galaxies we find 
an evolution of abundances (and element 
ratios) which is in good agreement with observations of DLAs 
over a large portion of the Hubble time, 
\item[$\bullet$] the models provide a smooth transition from DLAs  to 
nearby spirals' HII region abundances (Zaritsky \etal 1994),
\item[$\bullet$] the range of SF timescales 
$t_{\ast}$ as empirically derived for the near-by Hubble sequence of 
spiral galaxies from Sa through Sd fully explains the observed scatter of 
metallicities among DLAs at fixed 
redshift. 
\end{itemize}
(cf. Fritze - v. Alvensleben \& Fricke 1995, Fritze - v. Alvensleben 1995).

\begin{figure}
\epsfysize=10.0cm
\vskip -2.65cm
\centerline{\epsffile{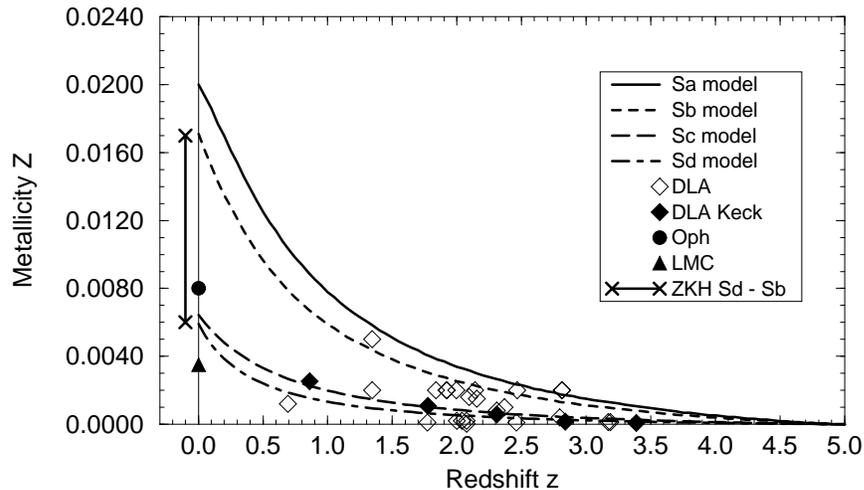}}
\vskip -1.0cm
\caption{Redshift evolution of the global metallicity: 
spiral models vs. DLA. 
ZKH: range of HII region abundances in nearby spirals of types Sb 
through Sd (Zaritsky {\sl et al.}). ${\rm (H_0,\Omega_0,z_{form})=(50,1.0,5)}$.}
\vskip -0.25cm
\end{figure}

\noindent
{\bf We conclude:} Our chemical evolution models confirm the hypothesis that 
DLA systems could be the progenitors of present day spirals and 
predict early as well as late type spirals to be among the DLA sample.

\noindent
In a complementary investigation Lindner \etal (1996, 1998) compare the 
spectrophotometric results from our 
unified chemical {\bf and} spectrophotometric model to a 
large sample of optically identified QSO absorbers. 

\section{Results from our Chemically Consistent Modelling}

In view of the low metallicities of DLAs a chemically consistent description is 
clearly desired. A problem with this approach is that important pieces of input 
physics, as e.g. stellar yields, explosion 
energies, lifetimes, mass loss rates, the final fate of a star and its remnant mass 
for very low metallicities are far from being fully understood. In this sense, 
the first results we now present from our chemically consistent modelling 
deserve some caution. 

\smallskip\noindent
Due to the complicated behaviour of the yields for individual elements as a 
function of stellar mass and metallicity, changes in the redshift 
evolution of chemically consistent models with respect to solar 
metallicity models vary for different elements, and, of course, also 
with galaxy type (=SFH). 

\smallskip\noindent
For Zn, the most common metallicity tracer in DLAs, our Sa -- Sd 
models nicely bracket 
recent and precise DLA abundance data over the redshift range 
${\rm 0.6 \lta z \lta 3.5}$ as seen in Fig. 2.

\begin{figure}
\epsfysize=8.5cm
\vskip -1.95cm
\centerline{\epsffile{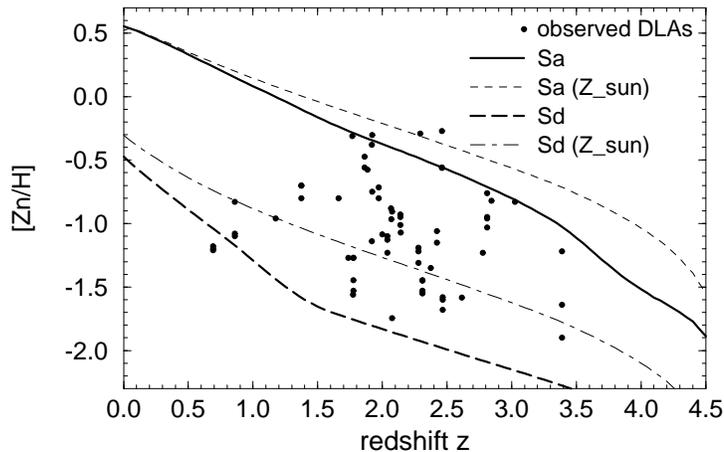}}
\vskip -0.7cm
\caption{Redshift evolution of [Zn/H] for spiral galaxy models 
together with DLA data.}
\vskip -0.3cm
\end{figure}

\smallskip\noindent
Abundance ratios change drastically as compared to solar 
metallicity models. This has far-reaching 
consequences e.g. for the interpretation of 
the stellar [Mg/Fe] in elliptical galaxies (see Fritze - v. Alvensleben 
1998). 

\smallskip\noindent
It is clearly desirable to extend this study 
to other elements with high resolution data. Our model offers a unique 
possibility to study the redshift evolution of the Global Cosmic SFR in the 
Universe including consistently all three observational accesses: gas content, 
chemical enrichment and spectral properties. 

\smallskip\noindent
{\sl UFvA greatfully acknowledges a travel grant (Fr 916/4-1) from the DFG.}


\begin{thebibliography}{}  

\bibitem[]{} Bahcall, J. N. \etal, 1993, ApJS {\bf 81}, 1
\bibitem[]{} Fritze - v. Alvensleben, U., 1995, in {\sl QSO Absorption Lines}, 
ESO Workshop, 
p. 81
\bibitem[]{} Fritze - v. Alvensleben, U., 1998, in {\sl Abundance Ratios of the Oldest Stars}, Highlights of Astronomy, {\sl in press}
\bibitem[]{} Fritze - v. Alvensleben, Fricke, K. J., 1995, IAU Symp. {\bf 164}, 457
\bibitem[]{} Fritze - v. Alvensleben, U., Kr\"uger, H., Fricke, K. J., 
Loose, H.-H., 1989, A\&A {\bf 224}, L1
\bibitem[]{} Fritze - v. Alvensleben, U., Kr\"uger, H., Fricke, K. J., 1989, 
A\&A {\bf 246}, L59
\bibitem[]{} Kennicutt, R. C., 1992, ApJS {\bf 79}, 255
\bibitem[]{} Kennicutt, R. C., Kent, S. M., 1983, ApJ {\bf 88}, 1094
\bibitem[]{} Matteucci, F., 1991, ASP Conf. Ser. {\bf 20}, 539
\bibitem[]{} Lindner, U., Fritze - v. Alvensleben, Fricke, K. J., 1996, A\&A {\bf 316}, 123
\bibitem[]{} Lindner, U., Fritze - v. Alvensleben, Fricke, K. J., 1997, in {\sl Structure and Evolution of the 
IGM from QSO Absorption Line Systems}, 
ed. Petitjean, 
{\sl in press}
\bibitem[]{} Lindner, U., Fritze - v. Alvensleben, Fricke, K. J., 1998, {\sl this volume}
\bibitem[]{} Lu, L., Sargent, W. L. W., Barlow, 
et al., 1996, ApJS {\bf 107}, 475
\bibitem[]{} Pettini, M., Smith, L. J., King, D. L., Hunstead, R. W., 1997, ApJ {\bf 486}, 665
\bibitem[]{} Stengler-Larrea, E., 1995, in {\sl QSO Absorption Lines}, 
ESO Workshop, 
p. 199
\bibitem[]{} van den Hoek, L. B., Groenewegen, M. A. T., A\&AS {\bf 123}, 305
\bibitem[]{} Woosley, S. E., Weaver, T. A., 1995, ApJS {\bf 101}, 181
\bibitem[]{} Zaritsky, D., Kennicutt, R. C., Huchra, J. P., 1994, ApJ {\bf 420}, 87


\end{thebibliography}
\end{document}